\documentclass[11pt]{article}

\usepackage{authblk}
\usepackage{fullpage}
\usepackage{amsmath}
\usepackage[version=3]{mhchem}

\usepackage{graphicx}
\usepackage{scicite}
\usepackage{times}
\usepackage{lineno}

\title{Applied Astrobiology:\\ An Integrated Approach to the Future of Life in Space}
\author[1]{R. Wordsworth} 
\author[1]{C. Cherubim} 
\author[2]{S. Nangle} 
\author[3,4]{A. Berliner} 
\author[5]{E. Dyson} 
\author[1]{P. Girguis} 
\author[6]{D. Grinspoon} 
\author[6]{R. Harris} 
\author[ ]{K. Liu} 
\author[7]{A. Marblestone} 
\author[8]{C. Mason} 
\author[9]{R. Morhard} 
\author[1]{D. Sasselov} 
\author[10]{S. Seager} 
\author[1]{R. Wood} 
\author[11]{P. Worden} 

\date{}

\affil[1]{Harvard University, Cambridge, Massachusetts, USA}
\affil[2]{Circe, Boston, Massachusetts, USA}
\affil[3]{Center for the Utilization of Biological Engineering in Space, Berkeley, California, USA}
\affil[4]{Weill Cornell Medicine, New York, New York State, USA}
\affil[5]{Wellville, New York, New York, USA}
\affil[6]{NASA, Washington, District of Columbia, USA}
\affil[7]{Convergent Research, Cambridge, Massachusetts, USA}
\affil[8]{Cornell University, Ithaca, New York, USA}
\affil[9]{Ginkgo Bioworks, Boston, Massachusetts, USA}
\affil[10]{Massachusetts Institute of Technology, Cambridge, Massachusetts, USA}
\affil[11]{Breakthrough Initiatives, Luxembourg}

\begin{document}

\maketitle


\section*{Abstract}
\emph{Searching for extraterrestrial life and supporting human life in space are traditionally regarded as separate challenges. However, there are significant benefits to an approach that treats them as different aspects of the same essential inquiry: How can we conceptualize life beyond our home planet?}

\section*{Introduction}

Astrobiology is a young field, having emerged in the 1990s alongside renewal of NASA's Mars exploration program and the discovery of the first exoplanets. In the last three decades, it has made significant progress, with major advances in our understanding of Earth's early evolution, the potential habitability of solar system planets, and the nature and diversity of exoplanets. Yet the fundamental objective of astrobiology -- discovering signs of life beyond Earth -- remains elusive. A central challenge is that, since Earth is the only known location in the universe that possesses life, astrobiology operates with a sample size of one (the ``$N = 1$'' problem).

During the same period, the space sector at large has seen accelerated growth. Since 1990, there have been over twenty five scientific missions launched to other planets by multiple countries, and the number of artificial satellites that orbit Earth has gone from under 500 to over 10,000 \cite{mason2024second}. While the human presence in space remains limited to low Earth orbit, this is likely to change in the coming years, with the NASA Artemis program and aims of SpaceX and Blue Origin to send humans to the Moon, Mars, and deep space.

Sustaining human life in space is a major technological challenge, and for long-term human missions to Mars and beyond, in situ resource utilization (ISRU) will be essential. Biotechnology, as represented by the emerging field of space bioprocess engineering, will play a pivotal role in ISRU, as it enables sustainable production of pharmaceuticals, food, and structural materials, and efficient reclamation of molecules. As the scale and duration of human missions increases and eventually transitions to permanent habitation, ecosystem-based approaches that involve many species and long-term adaptation to extraterrestrial environments will become increasingly important.

Given the developments in the fields of astrobiology and space bioprocess engineering, we see the search for and the expansion of life into space as increasingly aligned. Here we put forward the case for an expanded definition of astrobiology that encompasses the future of life in space. As we discuss below, the broad perspective that astrobiology provides has potential to accelerate technological progress, increase sustainability, and clarify the goals of sending humans and other forms of life to space. In turn, an applied approach to astrobiology will broaden our understanding of life itself, with the ultimate promise of resolving the central $N = 1$ problem of the field.

\section*{Near-Term Opportunities}

The ongoing expansion of the space sector is creating many new opportunities for astrobiology. As launch costs decrease, life detection experiments can be carried out more cheaply and more quickly than has been previously possible. One example is the Venus Life Finder Mission, which has a planned launch date of 2025 and aims to be the first privately funded mission to another planet \cite{seager2022venus}. Additional innovative astrobiology mission concepts include LIFE, a mission that would use interferometry to search for and characterize Earth-like planets around other stars \cite{quanz2022large}, a Mars-Venus-Earth human flyby mission on the SpaceX Starship for the 2030s \cite{mason2024second}, and the Nautilus Deep Space Observatory \cite{apai2022nautilus}, a fleet of inflatable space telescopes that would also use Starship. Existing astrobiology missions also stand to benefit from launch cost reductions. Mars Sample Return, NASA's flagship mission to bring martian rocks back to Earth in a search for signs of life, is currently under re-evaluation due to escalating costs, and involvement of the private sector may be part of the solution.

Additional opportunities will emerge once human missions to other planets become a reality. Mars, in particular, is a location where human explorers could transform the current robotic-based search for life. Human scientists working in situ possess an unparalleled ability to contextualize and process information from diverse sources. With sufficient mobility, robotic support, and laboratory capabilities, humans on the martian surface could conclusively answer the question of whether Mars once had life -- or possesses it today \cite{jakosky2024}.

Finally, experimental astrobiology will also be transformed by cheaper access to space. One of the major achievements of astrobiology over the last few decades has been to reveal the true range of extreme environments in which organisms on Earth can survive \cite{petersen2011hydrogen,colwell2013nature}. This has been accompanied by increasing understanding of the habitability of extraterrestrial environments. However, constructing exact laboratory analogs of the numerous stressors to life in these environments is extremely challenging. Detailed characterization of factors such as radiation damage, martian regolith toxicity, and acidity in venusian clouds requires new missions. Many additional insights can be gained by growing living organisms in space directly \cite{cottin2017space}, and this will also become much easier with increased access to space.

\section*{Astrobiology and Human Life Support}

Human life on Earth is possible only because we are sustained by, and a part of, the biosphere. Beyond Earth, we will likely need to bring an adapted biosphere with us. In the near term, biotechnology has a fundamental role to play in environmental control and life support systems \cite{nangle2020case,averesch2023microbial}. Bioprocess engineering has a number of advantages over more traditional approaches for life support, including scalability, flexibility, and low power consumption. Living systems are also uniquely effective at recycling key feedstocks and making efficient use of scarce resources \cite{santomartino2023toward}. Food production, material recycling and reclamation, and synthesis of pharmaceuticals and habitat materials are key areas where biology will provide significant practical and psychological benefits in the near future.

Despite the potential, key knowledge gaps remain regarding the role of biological life support in human missions. These gaps include the ability of bioreactor-based systems to perform reliably over long periods, the relative merits of different classes of organisms for specific roles, and the potential for synthetic biology to increase efficiency and produce useful materials. Initially, the deployment of biological life support systems in short-duration missions can be approached on a case-by-case basis. However, as the size of the human presence in space increases, so will the complexity and scale of life support, and a broader view will become increasingly important.
Because of its focus on life in space, as well as its close connections to planetary science and the study of habitable environments, astrobiology is a natural lens through which to view the long-term maintenance of human life in space. Astrobiology provides a deep understanding of how ecosystems and environments have co-evolved on Earth through time, which can serve as a basis for understanding how life can be sustained beyond Earth. Earth systems science emerged in the 1980s from early attempts to understand the co-evolution of Earth's climate and biosphere and has had major success as an organizing framework. Applied astrobiology has similar potential to enable the study of life in space as a general phenomenon.

One immediate way in which astrobiology can contribute to supporting human life in space is via the discovery and characterization of extremophiles and other exotic forms of life. Understanding the limits of life in extreme terrestrial environments may help identify new metabolisms for synthetic biology, as well as predict the expected performance of biosystems in unusual environments \cite{caro2024polyextremophile}. An astrobiological perspective can help spur new thinking on the capability of living organisms to create habitable environments \cite{wordsworth2024self}, as well as the potential for synthetic biology to achieve specific objectives in space, including biomining \cite{cockell2020space} and deep space exploration \cite{church2022picogram}.

\section*{Planetary Protection and Sustainability}

There is no historical precedent for extending life from Earth to other worlds, so the long-term consequences require careful consideration. Planetary protection aims to shield other worlds from terrestrial life and protect Earth from any extraterrestrial life, should it exist, until life detection missions have been conducted. Sustainability aims to allow human civilization to develop without irreversibly damaging the biosphere. In both cases, the overriding principle is that life and natural environments should be protected from serious harm. Given the similar motivations, an integrated approach will become essential once a permanent human presence exists beyond low Earth orbit. Here, again, astrobiology has a central role to play in the conversation.

Sustainability in space is a topic of growing importance \cite{virgili2016risk}. The most pressing immediate concern is pollution of low Earth orbit with space debris. As the space economy expands to the Moon and beyond, other issues such as how best to allocate and reuse scarce in situ resources will become increasingly important \cite{elvis2021concentrated}. Because of their effectiveness at recycling waste materials, biological systems will be essential to the development of sustainable settlements in space. Further research is required to understand the range of ways in which industrial and biological processes can usefully overlap, for example, the use of biology to reclaim industrial waste products \cite{berliner2021towards,santomartino2023toward}. The intense challenges posed by extraterrestrial sustainability are likely to lead to many innovations, with spin-off applications that help protect ecosystems back on Earth.

The goals of planetary protection are closely aligned to environmental and sustainability challenges. The current near absence of human spaceflight beyond low Earth orbit permits stringent protocols for the sterilization of spacecraft that operate at locations such as Mars, but as human exploration advances, a pragmatic approach inspired by best practice in sustainable development on Earth will be necessary. Given the likely future of human spaceflight, a window of opportunity exists in the coming years to perform life-detection missions at reduced cost, while planetary surfaces remain relatively undisturbed.

\section*{Robotics and Artificial Intelligence}

The rapid advances in AI and robotics over the last few years are opening up exciting new possibilities in space exploration. This topic intersects with astrobiology and the future of life in space in particularly interesting ways. In the broadest sense, we can regard robotics and machine intelligence as simply another manifestation of Earth's biosphere \cite{frank2022intelligence}. In this view, robotic space probes can be seen as part of the `extended phenotype' of the human species. On a practical level, however, there are still profound differences between biological systems and industrially manufactured machines.

Robots possess great potential to support life in space, given their unique ability to perform functions that are repetitive, delicate, or heavy-load bearing; also, they can operate in a vacuum and in other environments dangerous to humans. So far, most robots deployed beyond Earth have been produced in very small numbers by the intensive labor of large teams of highly skilled engineers. For many future applications, such as distributed in situ measurements across a planet's surface, telecommunications networks, or miniature deep-space probes, the balance is likely to shift to mass production, with greater tolerance for failure of individual units \cite{brooks1989fast}. Innovative new approaches such as bio-inspired robots capable of swarm behavior could have applications in diverse planetary environments (Ma et al. 2013). Soft, dexterous robots \cite{bartlett20153d} and machines constructed partly or entirely from renewable materials \cite{hartmann2021becoming} also have high potential for performing delicate tasks or operating in environments where feedstocks for manufacturing are scarce. Further ahead, partly or fully self-replicating robots \cite{dyson1979disturbing,moses2020robotic} remain far from practical implementation but have transformative potential.

Much more research is required to determine the ideal balance between biological systems and machines for aiding exploration and sustaining habitability beyond Earth in the future. As the complexity of the space economy and the capabilities of machines to perform beyond-human-level tasks continue to improve, the nature and roles of robots and AI systems in space will also evolve. An ecosystem-level astrobiological view will allow this question to be addressed in an integrated way.

\section*{Next Steps}

We propose a broadened definition of astrobiology that combines its traditional areas of focus with study of the future of life in space. Today, the human presence beyond Earth is tenuous. However, the rapid and accelerating growth of the space economy means that the time is right for basic research in this area.

There are a number of key areas where particular attention is needed to make progress. Most critically, dialogue between different sub-communities must increase. A cross-disciplinary view is necessary to integrate perspectives from systems biology, planetary science, robotics, and other fields, including the humanities. A broad approach that combines advancement of existing technologies with fundamental research is most likely to lead to new breakthroughs.

No single organization can effectively carry out this research program alone. The best path forward will involve a mix of publicly and privately funded research in academia, government, and the private sector. Because much of the technology discussed here is still far from commercial applications, privately funded Focused Research Organizations (FROs) may be particularly valuable for advancing technology in specific areas \cite{marblestone2022unblock}.

Given the long-term implications, this is an area where robust engagement from democratically controlled institutions and civil society is particularly critical. This could occur through direct research funding from government agencies such as NASA, non-profit field-building organizations, or other approaches. When possible, international collaborations should be prioritized.

As the most technologically advanced species on Earth, we have a crucial role in the biosphere, which includes the responsibility to protect existing ecosystems from irreparable damage. We also have a unique capability to explore worlds beyond Earth, and if we choose to, permanently spread life to them. Approached in the right way, expansion of life beyond Earth could allow us to reimagine our relationship with our own planet, while broadening our conception of what life elsewhere could be like. Ultimately, the $N = 1$ problem of astrobiology will be transformed once Earth is no longer the only known inhabited world.

Our planet is currently undergoing changes at an unprecedented pace. Seen in terms of Earth's entire history, the human spaceflight era is near-instantaneous -- if Earth's history were condensed into a 24-hour timespan, the period from 1961 to today would have a duration of 1.2 milliseconds. At this unique point in history, we have the opportunity as a species to decide what the future of life beyond Earth should look like. Once human expansion into space begins in earnest, the path will already be set, so the time for fundamental research in this area is now.

\bibliographystyle{abbrv}
\bibliography{ms}

\section{Acknowledgements}

This essay summarizes the outcome of a Radcliffe Exploratory Seminar held in June 2024. The authors thank the Radcliffe Institute for funding the workshop and creating an atmosphere conducive to stimulating discussion.

\end{document}